\documentstyle[twocolumn,epsf,seceq]{jpsj}
\def\nle{\ \raise.3ex\hbox{$<$}\kern-0.8em\lower.7ex\hbox{$\sim$}\ }
\def\nge{\ \raise.3ex\hbox{$>$}\kern-0.8em\lower.7ex\hbox{$\sim$}\ }

\def\chiapptau{\tilde{\chi}_{[T_2,T_1]}(\tau; t; t_{\rm w1})}
\def\chiappome{\tilde{\chi}_{[T_2,T_1]}(\tau_\omega; t; t_{\rm w1})}
\def\chish{\chi''_{[T_2,T_1]}(\omega; t; t_{\rm w1})}
\def\chit{\tilde{\chi}_{T(t)}(\tau_\omega; t)}
\def\chitorg{\chi''_{T(t)}(\omega; t)}
\def\CorisoT{C_{T}(\tau; t_{\rm w})}

\def\Corsh{C_{[T_2, T_1]}(\tau; t_{\rm w2}, t_{\rm w1})}
\def\Corshzer{C_{[T_2, T_1]}(\tau; 0, t_{\rm w1})}
\def\DelC{\Delta C_T(\tau; t_{\rm w})}
\def\DelCone{1 - C_T(\tau; t_{\rm w})}
\def\DelL{l_{\Delta T}}
\def\entshf{e_{[T_2, T_1]}(t; t_{\rm w1})}
\def\Ltshf{L_{[T_2, T_1]}(t_2, t_{\rm w1})}

\def\RtisoI{R_{T_1}(t_{\rm w1})}
\def\Rtshf{R_{[T_2, T_1]}(t; t_{\rm w1})}
\def\Tc{T_{\rm c}}

\def\tw{t_{\rm w}}
\def\tweff{t_{\rm w}^{\rm (eff)}}
\def\twone{t_{\rm w1}}
\def\twoneeff{t_{\rm w1}^{\rm (eff)}}
\def\twtwo{t_{\rm w2}}

\def\runtitle{Numerical Study on Aging Dynamics in Ising 
Spin-Glass Models: $T$-Change Protocols}
\def\runauthor{Tatsuo {\sc Komori}, Hajime {\sc Yoshino} and 
Hajime {\sc Takayama}}  

\title{Numerical Study on Aging Dynamics in Ising Spin-Glass Models:
  Temperature-Change Protocols}
\author{Tatsuo {\sc Komori}\footnote{Present address: Hydrographic 
Department, Maritime Safety Agency, 5-3-1 Tsukiji, Chuo-ku, Tokyo 
104-0045}, Hajime {\sc Yoshino}\footnote{E-mail:
 yhajime@ginnan.issp.u-tokyo.ac.jp} and Hajime 
{\sc Takayama}\footnote{E-mail: takayama@issp.u-tokyo.ac.jp}}
\inst{Institute for Solid State Physics, the University of Tokyo,
7-22-1 Roppongi, Minato-ku, Tokyo 106-8666}
\recdate{\today}
\inst{Institute for Solid State Physics, the University of Tokyo,
7-22-1 Roppongi, Minato-ku, Tokyo 106-8666}
\recdate{\today}

\abst{By means of Monte Carlo simulations on the three-dimensional Ising 
spin-glass model, we have studied aging phenomena with various 
temperature($T$)-change protocols.
Particularly, a $T$-shift protocol, in which a system is first quenched 
to and aged for a period $\twone$ at a temperature $T_1$, and subsequently 
aged at a new temperature $T_2$ is closely investigated. 
Most importantly, the mean size of domains, which is extracted from the 
replica-overlap function, was found to grow monotonically without any 
appreciable decrease by the $T$-change. We also found the relaxation of 
energy density and spin auto-correlation function during the $T$-shift
process can be explained within the following picture: 
the dynamics finally crossovers to an isothermal aging  of $T_2$ at
around $t_2 \simeq \twoneeff$ after the $T$-change, where $\twoneeff$
is an effective waiting time related with $\twone$, but in the
transient regime $t_2 \nle \twoneeff$, adjustment of the population of
thermally active droplets from that at $T_1$ to $T_2$ takes place very
slowly. Implications of the results on aging phenomena in other
$T$-change protocols such as $T$-cycling and continuous $T$-change
with an intermittent stop observed by simulations as well as
experiments are discussed.
}
\kword{spin glass, slow dynamics, aging, droplet theory}

\begin{document}
\sloppy
\maketitle

\section{Introduction}

Aging phenomena in spin glasses have been extensively studied in 
recent years. Experimentally, many elaborated protocols have been 
adopted to reveal various aspects of the 
phenomena.~\cite{VincHOBC,Nordblad,Weissman} Theoretically, on the 
other hand, many ideas and models on spin glasses and related 
systems have been proposed and examined to get a proper understanding 
of aging dynamics in such glassy systems.~\cite{BCKM, Bouchaud99}
On the side of numerical study, a number of simulations, which could 
account for qualitative features of the aging effects, have been 
performed.~\cite{Andersson,Rieger-94,Rieger-96} In spite of the
enormous efforts, however, a unified picture of aging dynamics in spin
glasses has not been established yet. The difficulty is directly
connected to nature of the low-temperature spin-glass (SG) phase which
has been a controversial issue since more than a decade ago.

Among simulational results on the short-ranged Ising SG models, of
particular importance is the growth of a mean domain size which is
observed through the replica-overlap function, $G(r,t)$, in an
isothermal aging
process.~\cite{Huse-91,Kisker-96,Marinari-98-VFDT,oursI}  In the
process a system is quenched instantaneously from temperature
above its SG transition temperature $\Tc$ to $T$ below $\Tc$, and is 
isothermally aged at $T$. The function $G(r,t)$ is defined as 
\begin{equation}
        G(r,t) = 
        \frac{1}{N} \sum_{i=1}^{N} \left[ 
        S_i^{(\alpha)}(t)S_i^{(\beta)}(t)
        S_{i+r}^{(\alpha)}(t)S_{i+r}^{(\beta)}(t) \right]_{\rm av},
\label{eq:repl-over}
\end{equation}
where $\alpha$ and $\beta$ are indices of two replicas which have
different random spin configurations at $t=0$ (an instant of the 
quench), and are independently updated. Its correlation length 
$R_T(t)$ is found to grow almost in a power-law in the case of 3
dimensional (3D) models,~\cite{Kisker-96,Marinari-98-VFDT,oursI} 
\begin{equation}
   R_T(t) \sim L_0(t/\tau_0)^{1/z(T)}, 
\label{eq:Rt-sim}
\end{equation} 
where $L_0$ and $\tau_0$ are certain characteristic units of length 
and time, respectively, and the exponent $1/z(T)$ depends on $T$. In our
previous work,~\cite{oursI} hereafter referred to as I, we obtained
\begin{equation}
   1/z(T) \simeq 0.17T, 
\label{eq:z(T)}
\end{equation} 
at $T < 0.7J$ for the Gaussian model with zero mean and variance of
interactions $J$. It is also fitted to a logarithmic functional form
written as $R_T(t) \simeq R_0 + b ({\rm ln}t)^{1/\psi}$ with $R_0, b$
and $\psi$ being constants.~\cite{Huse-91,Kisker-96,oursII}
This ambiguity is attributed to the insufficiency of the time window of 
simulations which can be performed by the presently available
computers. Although there remains such an ambiguity in the explicit
growth form of $R_T(t)$, we can naturally regard $R_T(t)$ as a
characteristic length scale of the SG ordering in aging dynamics,
i.e., a mean distance of domain walls separating different pure states. 

It is natural to ask to what extent the mean domain size $R_{T}(t)$ is 
related to aging behaviour observed in various quantities simulated. 
The droplet theory,~\cite{FH-86-PRL,BM-87-Heidel,FH-88-EQ,FH-88-NE} 
particularly the one proposed by Fisher and Huse,~\cite{FH-88-NE}
contains scaling ansatz which relate various quantities to the 
characteristic length scales such as $R_T(t)$. We have been
numerically studying aging phenomena in the 3D Ising SG model to 
clarify the validity of the ansatz.~\cite{oursI,oursII,oursIII}

\begin{figure}
\leavevmode\epsfxsize=55mm
\epsfbox{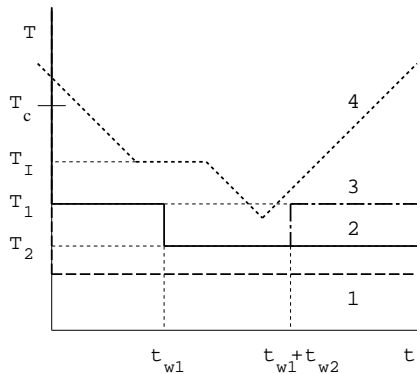}
\caption{Schedules of $T$-change in some experimental protocols of 
aging: 1) isothermal, 2) $T$-shift, 3) $T$-cycle, and 4) continuous 
$T$-change with an intermittent stop.
} 
\label{fig:protocol}
\end{figure}

Experimentally, various interesting phenomena have been also observed 
in aging processes in addition to the isothermal one. They are
$T$-shift, $T$-cycling, and continuous $T$-change with an intermittent
stop(s) protocols whose schedules of $T$-change are schematically
shown in Fig.~\ref{fig:protocol}. The ac susceptibility has been
continuously measured through these $T$-changes. In the case of a
negative $T$-shift process with $\Delta T \equiv T_2 - T_1 < 0$ in 
Fig.~\ref{fig:protocol} and with a magnitude $|\Delta T|$ larger than 
a certain value, the ac susceptibility is observed to increase 
discontinuously just after $t=\twone$ and then to relax 
again.~\cite{LHOV,AMN93,VincBHL} Its behaviour looks
as if the system restarts aging by the $T$-shift at $t=\twone$
forgetting its thermal history before $\twone$. It is called
the {\it rejuvenation}, or {\it chaos effect}. Furthermore when the 
temperature is turned back to $T_1$ at $t=\twone+\twtwo$ ($T$-cycling
protocol), the value of the ac susceptibility returns to that on the
isothermal aging curve of $T_1$ at almost $t=\twone$, and then relaxes
along the curve. This behaviour indicates that the system in fact 
remembers its history before $\twone$, and is called the {\it memory
effect}. Similar coexistence of the paradoxical phenomena, 
rejuvenation and memory effect, has been recently observed also in a 
continuous $T$-change with an intermittent stop(s).~\cite{JVHBN,JNVHB} 
These experimental findings have stimulated theoretical 
interest.~\cite{Bouchaud99}

The main purpose of the present paper is to investigate aging phenomena 
in various $T$-change protocols by simulations on the 3D Gaussian 
Ising SG model.~\cite{Komori-D} We concentrated mostly on the 
$T$-shift process, which is the most fundamental among the various 
$T$-change protocols, since the latter are considered as certain 
combinations of the former. We investigated the time evolution of the mean
domain size and the relaxation of the energy density and the spin
auto-correlation function during the $T$-shift process.

We have found, within the time window of our simulations, the following 
characteristic features of the aging dynamics. i) The mean
domain size $\Rtshf$ in the $T$-shift process (see 
Fig.~\ref{fig:protocol}), with either a negative or positive $\Delta T$,
does continue to grow. Here and hereafter $t$ is a time measured from
the first quench from above $\Tc$ to $T_1$, $t_2=t-\twone$ a period
that the system ages at $T=T_2$ after the $T$-shift at $t=\twone$. ii)
The time evolution of the energy density and the spin auto-correlation
function is related with the mean domain size $\Rtshf$ at sufficiently 
long time scales after the $T$-shift by the scaling relations which are 
the same as isothermal aging. iii) However, there is a transient time
regime after the $T$-shift where there is extra contribution to the
relaxations. Interestingly, the extra contribution also shows very
slow, non-exponential relaxation. We interpret the latter as not due
to the chaos effect predicted by the droplet 
theory~\cite{FH-88-EQ,FH-88-NE} but as due to adjustment of the
population of droplet excitations within each domain. 

The memory effect observed experimentally is easily interpreted from
the above-mentioned scenario, namely, it is attributed to the
persistence of domains in a $T$-shift process. It is not yet
conclusive, however, whether nature of the transient regime in the
scenario is common to phenomena observed experimentally just after a
$T$-shift. In this respect, the recent work on aging in the
ferromagnetic fine particles system (FFPS) done by Mamiya {\it et 
al.}~\cite{Mamiya99} is of quite interest, since the time window of
their observation in unit of the microscopic time ($\sim 10^{-3}$sec
for this system~\cite{MamiyaPr}) is rather closer to that of our
simulations. Indeed, they have observed $T$-shift aging phenomena
which can be well interpreted by our scenario. It seems, however, that
the scenario is hard to explain the rejuvenation, or chaos effect
observed experimentally in ordinary spin 
glasses,~\cite{LHOV,VincBHL,VincentPr,DjuJN} whose microscopic time is
of the order of $10^{-13}$sec. 

The present paper is organized as follows. In the next section we 
briefly review the scaling properties of isothermal aging derived 
from the droplet theory, add further comments on them, and present the
scenario on the $T$-shift process. In \S 3 we present the results of
our simulations on various aspects of the $T$-shift, as well as the
$T$-cycling and continuous $T$-change processes. In the final section 
implications of the results with the experimental observations are
discussed. 

\section{Phenomenological Picture}

\subsection{Scaling properties of isothermal aging}

Here we briefly review the scaling properties of isothermal aging 
derived by the droplet theory~\cite{FH-88-NE,FH-88-EQ} due to 
Fisher and Huse (FH). In an isothermal aging process a system is
quenched from above $\Tc$ to temperature $T$ below $\Tc$. At waiting
time $\tw$ after the quench, there are domain walls separating
different pure states of the SG phase at a typical distance
$R_T(\tw)$ from each other. The distance $R_T(\tw)$ can be
measured in simulations for instance through the replica overlap
function defined in eq.(\ref{eq:repl-over}).

Except for places where the domain walls run, the bulk of the system
is essentially in equilibrium. An important subtle feature inside the
domains of a spin glass is that there can be large scale thermal
fluctuations due to droplet excitations. In ideal equilibrium, a
droplet excitation can be defined as a global flip from a ground state
of a droplet (cluster) of spins within a distance $L/2$ from a certain
given site $i$.~\cite{FH-88-EQ} The latter can be considered as a
simple two-state system with a free-energy excitation gap $F_{L}(i)$
and a barrier energy $B_L(i)$ of a thermal activation process. 

The typical value $F^{\rm typ}_L$ of the gap $F_{L}(i)$ scales as, 
\begin{equation}
   F^{\rm typ} _L \sim  \Upsilon (L/L_0)^\theta,
\label{eq:FL-FH}
\end{equation}
where $\Upsilon$ is the stiffness constant of domain walls on the
boundary of the droplet. The gap is however broadly distributed  and
its distribution function $\rho_{L}(F)$ is considered to follow the  
scaling form,
\begin{equation}
       \rho_{L}(F) = {1 \over F^{\rm typ}_{L}}
       {\tilde \rho}\left( {F \over  F^{\rm typ}_{L}} \right). 
\label{eq:rho-FH}
\end{equation}
A very important property of the scaling function ${\tilde \rho}(x)$
is that it has finite intensity at $x=0$, ${\tilde \rho}(0)>0$, 
which allows large scale droplet excitations even at very low 
temperatures. The probability to have a thermally active droplet
at a very low temperature $T$ is given by 
\begin{equation}
p(L;T) \sim \frac{k_{\rm B}T}{\Upsilon  (L/L_{0})^{\theta}}\tilde{\rho}(0).
\label{eq:droplet-weight}
\end{equation}
The latter feature yields equilibrium properties which make
spin glasses distinctly different from simple
ferromagnets.~\cite{FH-88-EQ} 

The relaxation time of a droplet of size $L$ centered at site $i$ is
given by the Arrhenius law
\begin{equation}
\tau_{L}(i)=\tau_{0}\exp \left(\frac{B_{L}(i)}{k_{\rm B}T} \right),
\label{eq:arrhenius}
\end{equation}
where $B_L(i)$ is the energy barrier of the droplet. The typical value 
$B^{\rm typ}_L$ of the energy barrier $B_{L}(i)$ scales as,
\begin{equation}
B^{\rm typ}_L \sim \Delta (L/L_0)^\psi,
\label{eq:BL-FH}
\end{equation}
where $\Delta$ is a characteristic unit for energy barriers associated 
with thermal activation of droplet excitations, and the exponent $\psi$
satisfies $\theta \le \psi \le d-1$.

In the droplet theory, it is assumed that the growth of domains is
also governed by droplet excitations. An obvious but very important 
consequence of eqs.(\ref{eq:arrhenius}) and (\ref{eq:BL-FH}) is that
dynamical processes, including both the domain growth and droplet
excitations, at different length scales are extremely widely 
separated.~\cite{Bouchaud99} Hence within a time scale of small 
scale processes, large scale processes look as if they are almost
frozen.  

In isothermal aging, for example, the relaxation of the excessive energy
per spin $\delta e_{T}(t)$ with respect to the equilibrium value is
expected to follow the scaling form as, 
\begin{equation}
\delta e_{T}(t) 
\sim \tilde{\Upsilon}(R_T(t)/L_0)^\theta / (R_T(t)/L_0)^d, 
\label{eq:ene-sim}
\end{equation} 
where $\tilde{\Upsilon}$ is a characteristic energy scale, and $d$ is
the dimension of space which is 3 here. This scaling form was
confirmed, in our previous work I, with the energy exponent 
$\theta = 0.20 \pm 0.03$ which agrees with the result of the defect
energy analysis at $T=0$.~\cite{BM-84-theta} 

The droplet theory also predicts important scaling properties of the 
spin auto-correlation function,
\begin{equation}
        \CorisoT = \overline{ \langle S_i(\tau+\tw)S_i(\tw)\rangle }, 
\label{eq:def-Coriso}
\end{equation}
in isothermal aging. Here the over-line denotes the averages over 
sites and different realizations of interactions (samples), and the 
bracket denotes the average over thermal noises. In the so-called 
quasi-equilibrium regime defined by $\tau \ll \tw$, the typical size  
$L_T(\tau)$ of droplet excitations which take place in the time scale
of $\tau$ is much smaller than the typical separation $R_T(\tw)$ of
domain walls present after the waiting time $\tw$. For this situation the
droplet theory introduces a phenomenological concept called {\it
  effective stiffness} which characterizes a reduction of the excitation
gap of small scale droplets due to the presence of the domain
walls. The latter yields $\tw$-dependence of $\CorisoT$ which is
written in terms of $R_T(\tw)$ and $L_T(\tau)$ as the following:
\begin{equation}
\DelCone{} = \DelC{}+\alpha_T(\tau),
\label{eq:def-delc}
\end{equation}
with
\begin{equation}
\alpha_T(\tau) = 1-C_T^{\rm (eq)}(\tau),
\label{eq:alpha}
\end{equation}
and
\begin{equation}
\DelC{} \sim  \tilde{\rho}(0)
\frac{T}{\Upsilon}\left ( \frac{L_{0}}{L_T(\tau)}\right)^\theta
\left( \frac{L_T(\tau)}{R_T(\tw)} \right)^{d-\theta}.
\label{eq:delc}
\end{equation}

Then the zero field cooled susceptibility  
$\chi_{{\rm ZFC};T}(\tau; \tw)$ which is often observed by experiments
should also be understood within the same ansatz since the
fluctuation-dissipation theorem (FDT)  
\begin{equation}
\chi_{{\rm ZFC};T} (\tau; \tw) 
  \simeq {1 \over T}[1 - C_T(\tau; \tw)] 
  \equiv \tilde{\chi}_T(\tau; \tw). 
\label{eq:FDT-chi}
\end{equation}
holds in the quasi-equilibrium regime. The FDT has been tested and
confirmed by the careful experiments~\cite{AlbaHOR-87,RefregierOHV-88} 
even in the presence of weak violation of the time translational
invariance.

\subsection{Further comments on isothermal aging}

In our previous work,~\cite{oursII} hereafter referred to as II, we
confirmed the scaling ansatz on $\CorisoT$ described in the
previous subsection. Here, we first make the comment that 
$\chi_{{\rm ZFC};T}(\tau; \tw)$ experimentally observed and analyzed
through eq.(\ref{eq:FDT-chi}) follow the scaling ansatz.     

\begin{figure}
%\begin{center}
\leavevmode\epsfxsize=85mm
\epsfbox{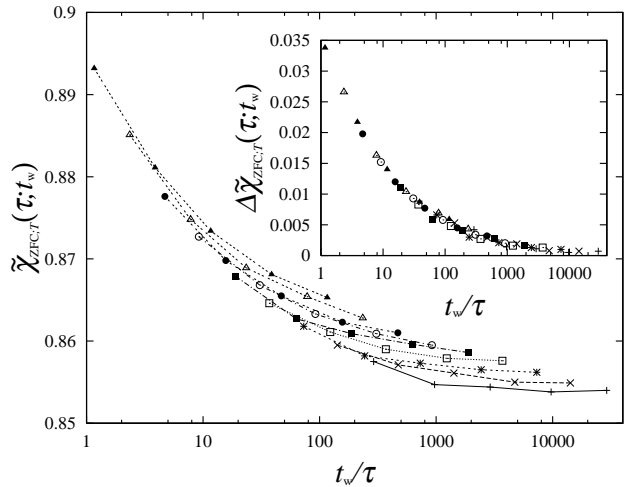}
\caption{$\chi_{{\rm ZFC};T}(\tau; \tw)$ of a AgMn spin
  glass~\cite{VincHOBC} at $T/\Tc=0.87$ for fixed $\tau$'s plotted
  against $\tw/\tau$. Only the data in the quasi-equilibrium regime
  are plotted, namely, they are for $\tau=1,\ 2,\ 4,\ 8,\ ...,\
  256$sec from bottom to top with $\tw=300,\ 1000,\ 3000,\ 10000,\
  30000$sec. In the inset the scaling behaviour of $\Delta\chi_{{\rm 
  ZFC};T}(\tau; \tw)$ corresponding to eq.(\ref{eq:delc}) is
  demonstrated.  
} 
\label{fig:m-meric}
\end{figure}

In Fig.~\ref{fig:m-meric}, for example, we plot 
$\chi_{{\rm ZFC};T}(\tau; \tw)$ of a Ag:Mn 2.6at\% spin glass against 
$\tw/\tau$. It is evaluated from $m_{{\rm TRM};T}(\tau; \tw)$
measured by the Saclay group~\cite{VincHOBC} as 
$\chi_{{\rm ZFC};T}(\tau; \tw) = [m_{{\rm FC};T} - 
m_{{\rm TRM};T}(\tau; \tw)]/h$ with $m_{{\rm FC};T}$ being the
field-cooled magnetization under a field of strength $h$. We note that 
we plot here only a part of their data with $\tw\nge\tau$, i.e., those 
in the quasi-equilibrium regime.
The data for different $\tau$'s lie on a universal curve when they are 
vertically shifted by suitable amounts depending on $\tau$ as shown in 
the inset of the figure. The feature is quite similar to 
$1 - C_T(\tau; \tw)$ analyzed in II. [Note that the factor 
$(L_0/L_T(\tau))^\theta$ is practically constant in eq.(\ref{eq:delc}) 
since $\theta$ is much smaller than $d-\theta$, as was the case of 
$1 - C_T(\tau; \tw)$ in II.] This indicates that 
$\chi_{{\rm ZFC};T}(\tau; \tw)$, via eq.(\ref{eq:FDT-chi}), follows
the scaling forms eqs.(\ref{eq:def-delc}), (\ref{eq:alpha}) and 
(\ref{eq:delc}) as well. The values of the exponents $(d-\theta)/z(T)$ 
and $\theta/z(T)$ turn out to roughly agree with the estimates
obtained by making use of our previous results in II, $\theta \simeq
0.20$ and $1/z(T) \simeq 0.16(T/\Tc)$, i.e., eq.(\ref{eq:z(T)}) with 
$\Tc=0.95\pm 0.04$,~\cite{Marinari-cd98-PS} However, such
semi-quantitative agreement is lost for the data at temperatures
closer to $\Tc$. 

An intriguing point is that the above result suggests that $R_T(t)$ in
the time scale of experiments is also compatible with the estimate
from our simulations. In this context it is worth pointing out that 
Joh {\it et al.}~\cite{Joh-98} have recently extracted $R_T(t)$ of
Cu:Mn 6at\% and CdCr$_{1.7}$In$_{0.3}$S$_4$ spin glasses from the
thermoremanent magnetization (TRM). The growth law they have obtained
turns out to agree with the simulational result of 
eqs.(\ref{eq:Rt-sim}) and (\ref{eq:z(T)}) even quantitatively. The
estimated $R_T(t)$ with $t$ of the order of laboratory time is at most
only several tens of lattice distances. However, the experimental
studies on the isothermal aging dynamics so far reported have not
converged to a unified picture yet as we discussed in II.

Another comment is related to the characteristic length and time
scales, $L_0$ and $\tau_0$ in eq.(\ref{eq:Rt-sim}), which we did not
fix explicitly in II. According to the droplet theory~\cite{FH-88-EQ},
they are some microscopic scales at lower temperatures, but near below
$\Tc$ they are expected to crossover to the SG coherence length
$\xi_-$ and the corresponding critical relaxation time,
respectively. Indeed, Hukushima {\it et al.}~\cite{Huku} have found
recently such a crossover behaviour in aging dynamics in the 4D $\pm
J$ Ising SG model. A similar analysis on the present 3D model
certainly has to be done, which may clarify the disagreement between
the experimental and simulational results close to $\Tc$ mentioned
just above. Furthermore, as suggested in Ref. 34, it may also
provide a useful key to resolve the apparent ambiguity of the growth
law mentioned previously. 

\subsection{$T$-shift process within overlap-length}

Now let us consider the $T$-shift process. After a system is first 
quenched from above $\Tc$ to a temperature $T_1$ below $\Tc$, domains
of different pure states grow up. Their mean size reaches to $\RtisoI$
after a given waiting time $t=\twone$. After the temperature is changed at 
$t=\twone$ to a new temperature $T_2$, we expect the domains continue
to grow but with a rate specific to the new temperature $T_2$ (see 
eq.(\ref{eq:tc-xi}) below). It can be said that, at the instant of the 
$T$-shift, the system is aged as if it is aged for the effective waiting
time $\twoneeff$ by isothermal aging at $T_{2}$, where $\twoneeff$
is determined by
\begin{equation}
        R_{T_1}(\twone) \simeq R_{T_2}(\twoneeff),
        \label{eq:tweff}
\end{equation}
with $R_{T}(t)$ being the growth law of the mean domain size in
isothermal aging at $T$.

Another important process which must take place after the $T$-shift is 
adjustment of the population of thermally active droplets within the
domain. Since the equilibrium population is proportional to $T$ as
given in eq.(\ref{eq:droplet-weight}), there must be a change of the
population of an amount proportional to $\Delta T=T_{2}-T_{1}$ at each 
length scale $L$. It is emphasized that this process itself is
intrinsically very slow since adjustment of the droplets takes place
only by activation processes described by eqs.(\ref{eq:arrhenius}) and 
(\ref{eq:BL-FH}). Thus we naturally expect that there is a new
characteristic length $\Ltshf$ which slowly grows after the $T$-shift
such that the population of droplets smaller than $\Ltshf$ is
equilibrated to the new temperature $T_2$. However, the population of 
droplets larger than $\Ltshf$ is still equilibrated with respect to
the old temperature $T_{1}$. We call the characteristic length
$\Ltshf$ a size of {\it quasi-domains}. 

The growth law of $\Ltshf$ is expected to be the same as the domain. 
When the system ages at $T_2$ for a period about $\twoneeff$ defined by 
eq.(\ref{eq:tweff}), $\Ltshf$ catches up $\Rtshf$ and the
quasi-domains merge into domains in isothermal aging at $T_2$. Thus at 
large time scales $t_2 \nge \twoneeff$, aging becomes essentially the
same as isothermal aging at $T_{2}$. The time range 
$t_2 \nle \twoneeff$, on the other hand, is regarded as a transient
regime where adjustment of the population of droplets are taking place.

In the above argument we implicitly assumed that equilibrium states
at different temperatures are the same except for the change of the
population of droplets eq.(\ref{eq:droplet-weight}). It should be
remarked that the assumption makes sense only for length scales
smaller than the so-called overlap length 
$\DelL$.~\cite{FH-88-EQ,BM-87-chaos} 
It is predicted that equilibrium states at different nearby 
temperatures $T_1$ and $T_2$ differ from each other at length scales
larger than this length $\DelL$. This aspect is called {\it chaotic}
nature of the SG phase.~\cite{BM-87-chaos}  The overlap length $\DelL$
is expected to diverge when $\Delta T = T_2 - T_1 \rightarrow 0$ as
\begin{equation}
        \DelL \sim L_0 \left( {\Upsilon^{3/2} \over T^{1/2} |\Delta T|} 
       \right)^{2/(d_s-2\theta)},
\label{eq:overlapL}
\end{equation}
where, $\theta$ is the exponent in eq.(\ref{eq:FL-FH}) and $d_s$ the
fractal dimension of surface of the droplets. 

\section{Results of Simulations}

\subsection{Model and method}

We have carried out MC simulations on various aging processes 
in the same 3D Ising SG model as in our previous works, i.e., 
the one with Gaussian nearest-neighbor interactions with mean zero and 
variance $J=1$. The critical temperature has been obtained as 
$\Tc=0.95\pm 0.04$.~\cite{Marinari-cd98-PS}. 
The heat-bath MC method we use here is also the same as in I and II. 
The data we will discuss below are obtained mostly at $T=0.6 \sim
0.8$ in systems with linear size $L_{\rm s}=24$ averaged over a few
hundreds samples with one MC run for each sample. In I, it was
confirmed that finite size effects do not appear for these values of 
$T$ and $L_{\rm s}$ within our time window ($\nle 2 \times 10^{5}$
MCS). 

\subsection{SG coherence length --- domain growth}

\begin{figure}
\leavevmode\epsfxsize=80mm
\epsfbox%
{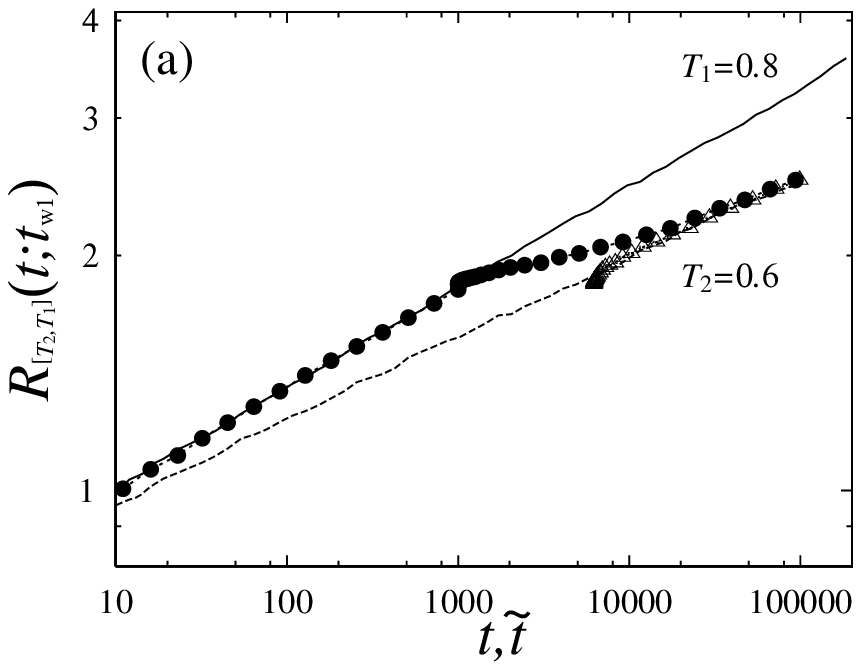}

\leavevmode\epsfxsize=80mm
\epsfbox%
{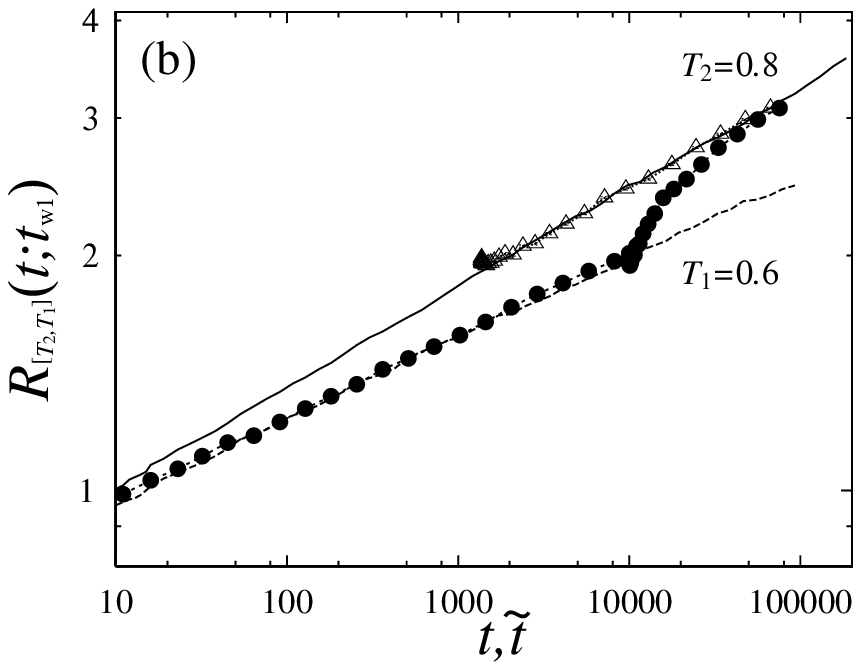}
\caption{(a) Growth of $\Rtshf{}$ in a negative $T$-shift process 
with $T_1=0.8, \ T_2=0.6$ and $\twone=1000$ MCS (solid circles). 
The solid and broken lines are the isothermal aging curves 
$R_{T_1}(t)$ and $R_{T_2}(t)$, respectively. The open triangles represent 
$\Rtshf{}$ plotted against $\tilde{t} = t_2 + \twoneeff$ where 
$t_2 = t - \twone$ and $\twoneeff = 6230$. (b) $\Rtshf{}$ in a 
positive $T$-shift process with $T_1=0.6, \ T_2=0.8$ and 
$\twone=10000$ MCS. The symbols and curves are the same as those in 
(a) but with $\twoneeff = 1400$. 
}
\label{fig:tc-xi}
\end{figure}

Let us begin with time evolution of $\Rtshf{}$, a mean domain size in 
a negative $T$-shift protocol with $\Delta T = T_2 -T_1 < 0$ (see 
Fig.~\ref{fig:protocol}). The simulated results with 
$T_1=0.8,\ T_2=0.6$ and $\twone=1000$ MCS are presented by solid 
circles in Fig.~\ref{fig:tc-xi}(a). They are extracted from the 
replica overlap function eq.(\ref{eq:repl-over}) which is obtained 
by averaging over several thousands samples with $L_s=32$.

Clearly, $\Rtshf{}$ does grow continuously without any noticeable 
decrease by the $T$-shift. This implies that, just after the
$T$-shift, the system looks as if it has aged isothermally at $T=T_2$
for a period of the effective waiting time $\twoneeff$ defined in 
eq.(\ref{eq:tweff}). Indeed, when $\Rtshf{}$ at $t>\twone$ is plotted
against $\tilde{t}=\twoneeff+t_2$ (open triangles in the figure), it
almost lies on the isothermal aging curve $R_{T_2}(t)$, i.e., the two 
curves are related as
\begin{equation}
        \Rtshf \simeq R_{T_2}(t_2+\twoneeff),
\ \ {\rm at} \ \ t_2 \nge 0.
        \label{eq:tc-xi}
\end{equation}
Here, a value of $\twoneeff \ (\simeq 6230)$ has been chosen so as to
obtain a good collapse of the two curves at time range 
$t_2 \nge \twoneeff$. The chosen value turns out to satisfy 
eq.(\ref{eq:tweff}) combined with eqs.(\ref{eq:Rt-sim}) and
(\ref{eq:z(T)}).  

As shown in Fig.~\ref{fig:tc-xi}(b), $\Rtshf{}$ in a positive 
$T$-shift protocol is also described by eq.(\ref{eq:tc-xi}), but 
with $\twoneeff \ (\simeq 1400) < \twone\ (=10^4)$ in this case. The 
continuous growth of domains in the $T$-shift process, either a
negative or positive one, is one of the most important results 
observed in the present simulation.

\subsection{Evolution of energy --- Transient and isothermal regimes}

\begin{figure}
\leavevmode\epsfxsize=85mm
\epsfbox%
{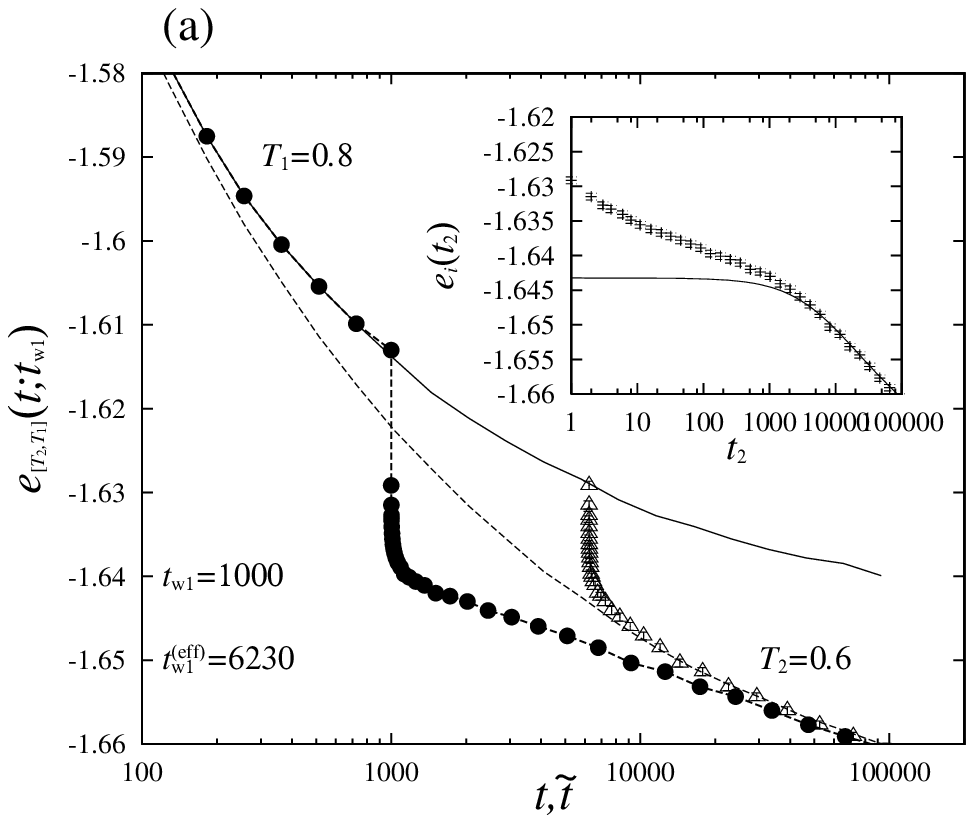}

\leavevmode\epsfxsize=85mm
\epsfbox%
{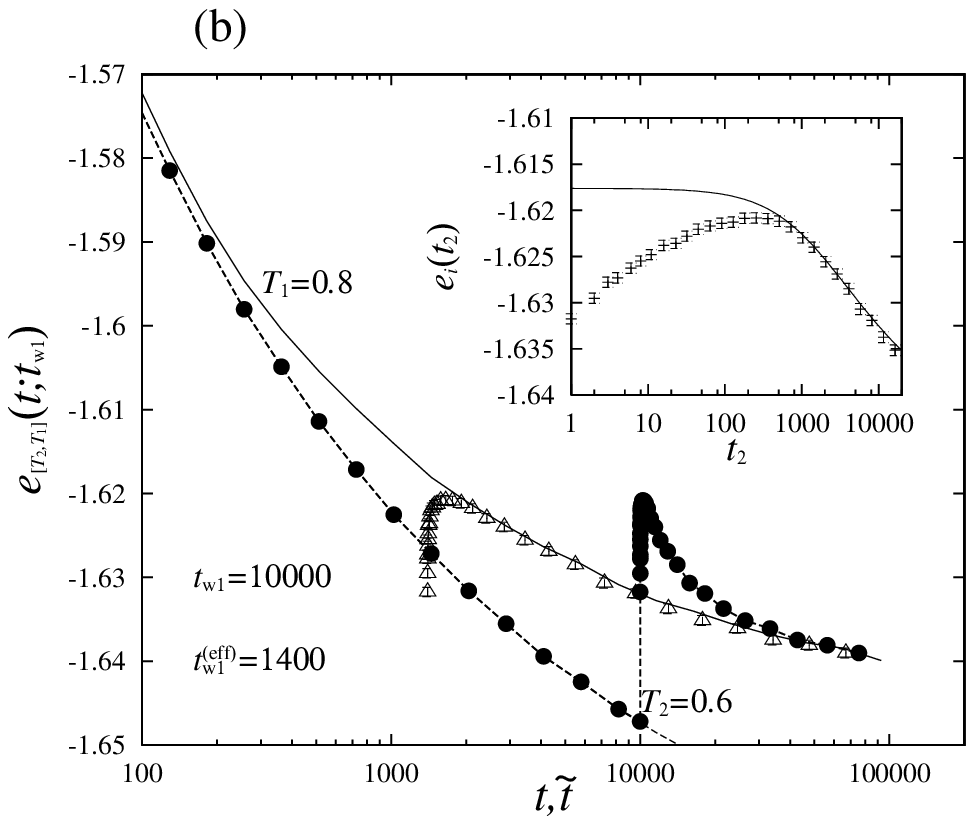}
\caption{Time evolution of the energy per density, $\entshf{}$, in the 
same (a) negative and (b) positive $T$-shift protocols as in 
Fig.~\ref{fig:tc-xi}. The symbols and curves represent 
$\entshf{}$ and $e_T(t)$ in replacement of $\Rtshf{}$ and $R_T(t)$, 
respectively. In the insets of both figures $\entshf{}$ as a function
of $t_2$ and $e_{T_2}(t-\twoneeff)$ are shown by the data points and the
curve, respectively. 
}
\label{fig:tc-ene}
\end{figure}

In Fig.~\ref{fig:tc-ene}(a) we present the time evolution of energy per 
spin, $\entshf{}$, in the same negative $T$-shift process as shown in 
Fig.~\ref{fig:tc-xi}(a). The original data drawn by the solid circles 
quickly deviate from the isothermal aging curve, $e_{T_1}(t)$, just 
after the $T$-shift at $t=\twone$, and then slowly crossover to 
$e_{T_2}(t)$ from below at longer times. As drawn by the open triangles 
in the figure, the part of $\entshf{}$ after the crossover can be 
laid upon $e_{T_2}(t)$ if it is shifted rightwards properly, i.e., 
the two curves are related as
\begin{equation}
        \entshf \simeq e_{T_2}(t_2+\twoneeff)
\ \ \ {\rm at} \ \ t_2 \nge \twoneeff,
        \label{eq:tc-ene}
\end{equation}
with the same $\twoneeff$ as the one determined before through
$\Rtshf{}$.
Notice that the shifted $\entshf{}$ approaches $e_{T_2}(t)$ from 
above. 

In the positive $T$-shift process, $\entshf{}$ behaves 
similarly and eq.(\ref{eq:tc-ene}) also holds as shown in 
Fig.~\ref{fig:tc-ene}(b). In this case the original (shifted) 
$\entshf{}$ merges into $e_{T_2}(t)$ from above (below).

From inspection of Figs.~\ref{fig:tc-xi} and \ref{fig:tc-ene} we may
divide the aging process after the $T$-shift into the following three
regimes. 
\begin{enumerate}
\renewcommand{\labelenumi}{\arabic{enumi})}
\item Quasi-reversal regime ($t_2 \le t_{\rm rev} \simeq 1$)\\
Each spin responds individually to changes in its local Boltzmann 
weights. This gives rise to a relatively large change in $\entshf{}$ 
in the first MC step after the $T$-shift. The process is expected 
almost reversible, though we have not explicitly confirmed it.
\item Transient regime ($t_{\rm rev} \nle t_2 \nle \twoneeff$)\\
Judging from the behaviour of $\Rtshf{}$, eq.(\ref{eq:tc-xi}), we 
may regard that the system has aged roughly for $\twoneeff$ at $T=T_2$ 
already at around $t_2 \sim t_{\rm rev}$. Actually, $\entshf{}$ is 
significantly smaller (larger) than $e_{T_2}(t)$ at $t \nge \twone$ 
in the negative (positive) $T$-shift process.\\ 
\hspace*{3mm}
Another interesting comparison is between $\entshf{}$ as 
a function of $t_2$ and $e_{T_2}(t-\twoneeff)$ which is shown in the 
insets of Fig.~\ref{fig:tc-ene}. Now the former is larger (smaller) 
than the latter in the negative (positive) $T$-shift process, and it 
relaxes to the latter slowly, by no means exponentially. 
Within our scenario introduced in \S 2.3, these can be
explained qualitatively as follows. Just after the negative (positive) 
$T$-shift, the population of active droplets within the domain
is larger (smaller) than in equilibrium at $T_2$. Thus the energy
density decreases (increases) toward the isothermal curves.
\item Isothermal regime ($t_2 \nge \twoneeff$)\\
In this regime time evolution of the system is nothing but isothermal 
aging at $T=T_2$.  The result eq.(\ref{eq:tc-ene}) combined with 
eq.(\ref{eq:tc-xi}) means that the mean domain size and the energy
density are related by the scaling relation eq.(\ref{eq:ene-sim}).
The latter implies that relaxation of the energy density in this regime
is due to the coarsening of domain walls.
\end{enumerate}

The existence of isothermal regime 3), to which the system crossovers 
from regime 2) at around $t_2 \sim \twoneeff$, is another important 
aspect of the aging dynamics observed by the present simulation. It is 
certainly a consequence of the persistence of domains through the 
$T$-shift process described before, and can be regarded as one of the 
memory effects. An interesting question is if the phenomena in
transient regime 2) are related with the rejuvenation, or chaos effect 
observed experimentally. 

\subsection{Spin auto-correlation function}

We define the spin auto-correlation function in the $T$-shift process as
\begin{equation}
     \Corsh{} = \overline{ 
    \langle S_i(\tau+\twone+\twtwo)S_i(\twone+\twtwo)\rangle }.
\label{eq:def-Cor}
\end{equation}
and measure relaxation of 
\begin{equation}
     \chiapptau \equiv {1 \over T}[1-C_{[T_2,T_1]}(\tau;
     t_2,\twone)]
\end{equation}
with increasing $t_{2}$ after the $T$-shift. At each measurement, a
fixed value of the time separation $\tau$ is chosen as a parameter. 

We naively expect that the FDT holds for time regime $\tau < \twtwo$
as in the case of isothermal aging (see eq.(\ref{eq:FDT-chi})). If it
is the case the above $\chiapptau$ is identical to the ZFC
susceptibility. In experiments, relaxation of the out-of-phase
component of the ac susceptibility with a fixed frequency, say $\omega$, 
is most frequently measured also in the $T$-shift protocol. 
It has been numerically ascertained in II that the relaxation of
the ac susceptibility at a fixed frequency $\omega$ is very 
similar to that of the ZFC susceptibility with a fixed 
$\tau=2\pi/\omega\equiv \tau_\omega$. Thus, in the present work, we
regard that $\chiappome$ is essentially equivalent to the ac
susceptibility measured in experiments. 

The function with $\twtwo=0$, denoted by $\Corshzer$,  was 
examined in our previous work\cite{oursIII} hereafter referred to as 
III. We found that at $\tau \ll \twoneeff$ it follows similar scaling 
forms to those for $C_T(\tau;\tw)$ in isothermal aging described in 
\S 2.1 if the factor $L_T(\tau)/R_T(\tw)$ in eq.(\ref{eq:delc}) is 
replaced by $L_{[T_2,T_1]}(\tau;\twone)/R_{T_1}(\twone)$. Here 
$L_{[T_2,T_1]}(\tau;\twone)$ is a mean size of the droplet 
excitations, or quasi-domains, which are in local equilibrium at 
the new temperature $T_2$ within time scale $\tau$ after the
$T$-shift.  

\begin{figure}
\leavevmode\epsfxsize=80mm
\epsfbox%
{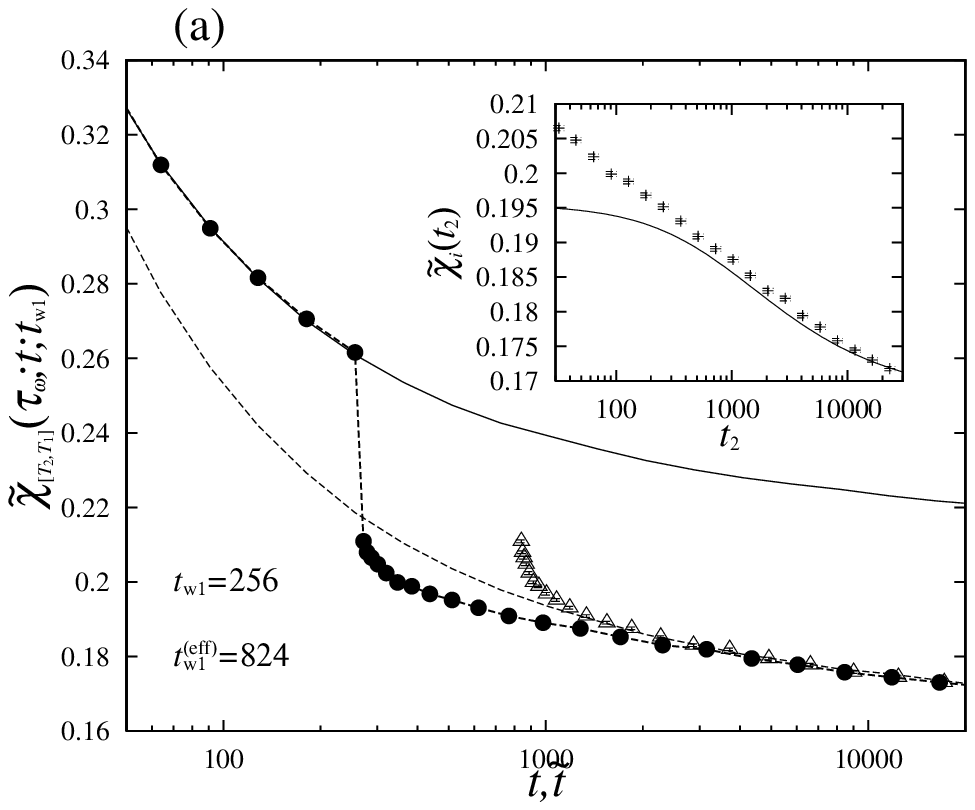}

\leavevmode\epsfxsize=80mm
\epsfbox%
{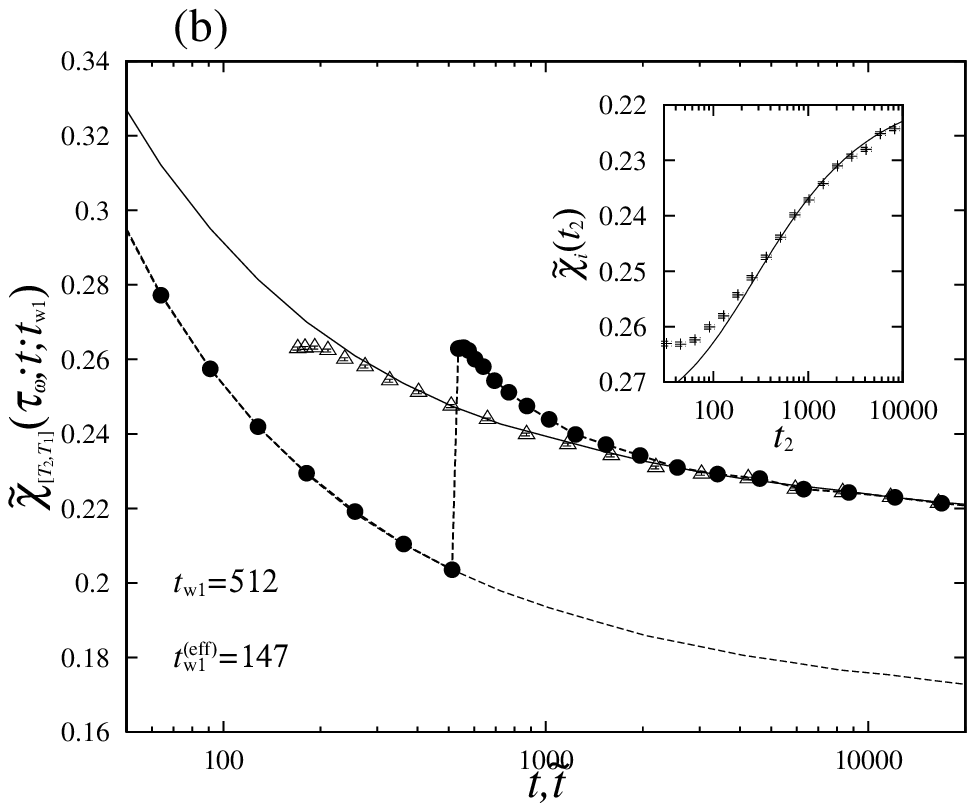}
\caption{$\chiappome$ in a negative $T$-shift with $T_1=0.8,
\ T_2=0.6$ (a) and in a positive $T$-shift with $T_1=0.6,
\ T_2=0.8$ (b). The symbols and curves, including those in the 
insets, represent  $\chiappome$ and $\tilde{\chi}_T(\tau_\omega;t)$
in replacement of $\entshf{}$ and $e_T(t)$ in Fig.~\ref{fig:tc-ene}), 
respectively. The values of $\twone$ and $\twoneeff$ are denoted in
each figure. 
}
\label{fig:chiappro}
\end{figure}

In Fig.~\ref{fig:chiappro} we show typical results of $\chiappome$ 
simulated for a fixed $\tau_\omega\ (=32)$. Their behaviour is
rather similar to that of $\entshf$ shown in Fig.~\ref{fig:tc-ene}. 
Just after the $T$-shift, it rapidly decreases (increases) and then 
relaxes slowly to the isothermal curve 
$\tilde{\chi}_{T_2}(\tau_\omega; t)$ from below (above) for 
$\Delta T < 0\ (>0)$. Furthermore, it also satisfies a relation
similar to eq.(\ref{eq:tc-ene}): when the part of its curve at $t_2>0$
is plotted against $\tilde{t}= t_2 + \twoneeff$ it now relaxes to the 
isothermal curve from above (below), 
\begin{equation}
        \chiappome \simeq \tilde{\chi}_{T_2}(\tau_\omega; t_2+\twoneeff)
\ \ \ {\rm at} \ \ t_2 \nge \twoneeff.
        \label{eq:tc-chi}
\end{equation}
Here, a value of 
$\twoneeff \ (\simeq 824)$ has been chosen so as to obtain a good 
collapse of the two curves at the time range $t_2 \nge \twoneeff$. The
resultant $\twoneeff$ is again consistent with eq.(\ref{eq:tweff})
combined with eqs.(\ref{eq:Rt-sim}) and (\ref{eq:z(T)}) obtained in I.

The above feature in isothermal regime 3), $t_2 \nge \twoneeff$, means 
that there the mean domain size and the correlation function are
related by the scaling relations of eqs.(\ref{eq:def-delc}),
(\ref{eq:alpha}) and (\ref{eq:delc}). Thus relaxation of $\chiappome$
in this regime is due to increase of the effective stiffness due to the
coarsening of domain walls.

Behaviour of $\chiappome$ in transient regime 2) is considered to 
reflect nature associated with quasi-domains. Let us here focus on 
the negative $T$-shift shown in Fig.~\ref{fig:chiappro}(a). As pointed
out just above, $\chiappome$ is definitely smaller than 
$\tilde{\chi}_{T_2}(\tau_\omega;t)$ at $t \nge \twone$. This is
different from the rejuvenation, or chaos effect observed in
experiments if the latter means that the system after the $T$-shift
looks almost as young as the system which has just started a new
isothermal aging at $T=T_2$. If this is the case, $\chiappome$ as a
function of $\tilde{t}$ with $\twoneeff$ {\it smaller} than $\twone$
should coincide with $\tilde{\chi}_{T_2}(\tau_\omega;t)$ at 
$t_2 \nge t_{\rm rev}$.~\cite{Mamiya99} 
The behaviour of $\chiappome$ may suggest that, if the overlap length
$\DelL$ introduced by eq.(\ref{eq:overlapL}) in \S 2.3 exists in the
present SG model, it should be larger than $R_{T_1}(\twone)$ observed
in the simulation. 
 
Let us then compare $\chiappome$ as a function of $t_2$ with 
$\tilde{\chi}_{T_2}(\tau_\omega;t-\twoneeff)$. The two curves are 
shown in the inset of Fig.~\ref{fig:chiappro}(a). Interestingly, the
former is larger than the latter and the excessive part decays very
slowly. This feature in transient regime 2) may be explained within
the scenario sketched in \S 2.3 as the following. After $\twtwo$ from
the $T$-shift, there remains excessive population of droplets of
length scales larger than the size of the quasi-domain $\Ltshf$. The
excessive population is proportional to $\Delta T=T_{2}-T_{1}$. The
boundary of the latter droplets may act as frozen-in domain walls and
reduces the effective stiffness of droplets smaller than $\Ltshf$ just
as in the case of the domain walls which separates different pure
states.~\cite{FH-88-NE,oursII} This yields the extra contribution,
which is proportional to $\Delta T$, to the relaxation. The scaling
analysis on the excessive part in terms of $\Ltshf$ and $\Rtshf{}$ is
now under investigation. 

\subsection{$T$-cycling process}

\begin{figure}
\leavevmode\epsfxsize=80mm
\epsfbox%
{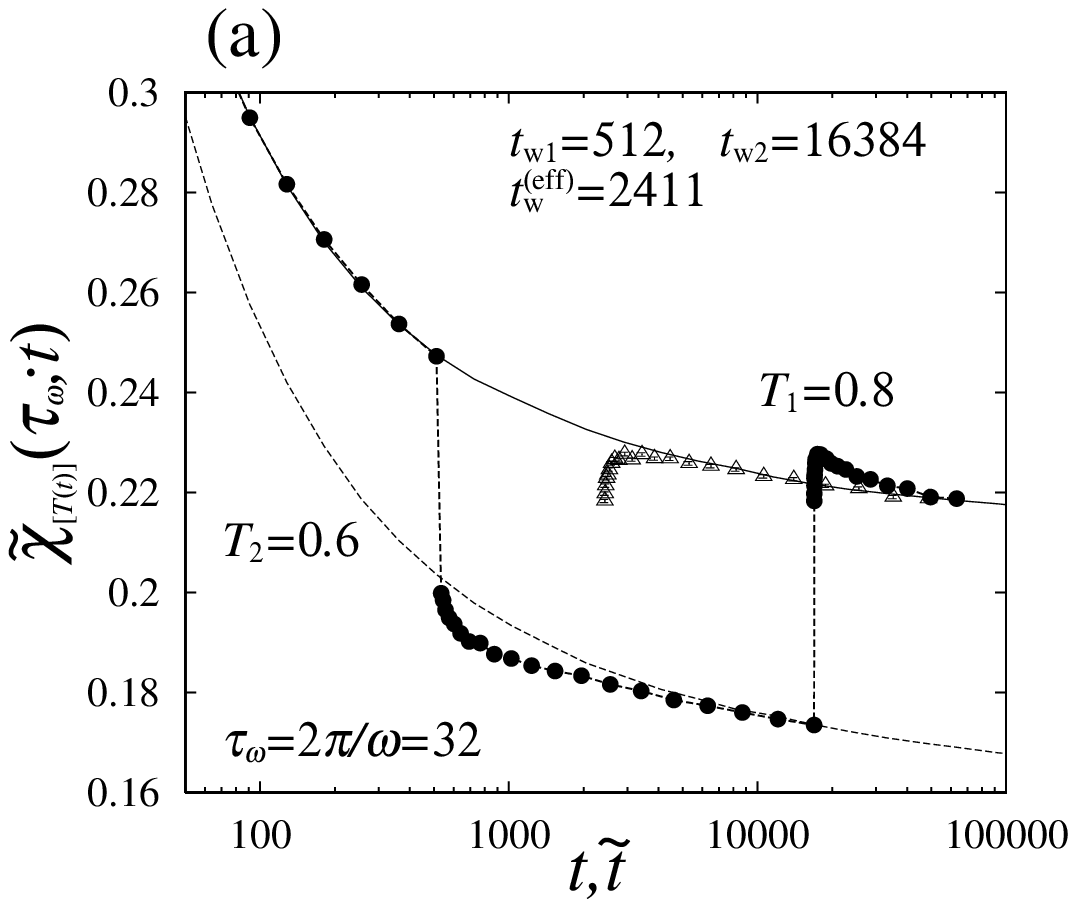}

\leavevmode\epsfxsize=80mm
\epsfbox%
{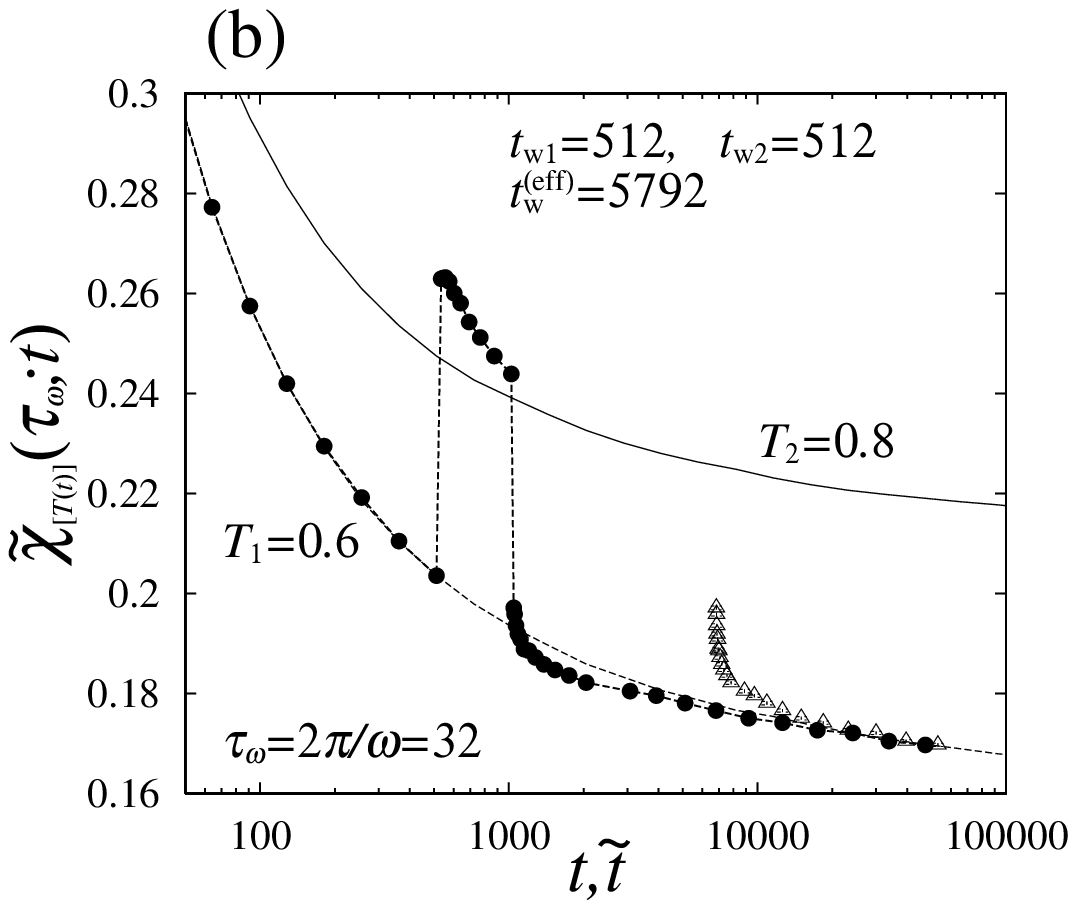}
\caption{$\chit$ in a negative $T$-cycling with $T_1=T_3=0.8,\ 
T_2=0.6$ (a) and in a positive $T$-cycling with $T_1=T_3=0.6,\ 
T_2=0.8$ (b). The symbols and lines are the same as those in 
Fig.~\ref{fig:chiappro} but with 
$\tilde{t} = t_3 + t_{\rm w}^{\rm (eff)}$.
}
\label{fig:Tcyc}
\end{figure}

A $T$-cycling process is a combination of two $T$-shift processes 
with the same $|\Delta T|$ but with opposite signs. At a time 
$t_2=\twtwo$ after the $T$-shift from $T_1$ to $T_2$ the temperature is 
turned back to $T_1$, and the subsequent relaxation is observed. When 
$\twtwo \gg \twoneeff$, the $T$-change back to $T_1$ is effectively
the same as a simple $T$-shift process from an isothermally aged state
at $T_2$ to $T_1$, as shown in Fig.~\ref{fig:Tcyc}(a). 
In this negative $T$-cycling process ($T_2<T_1$), the approximated ac 
susceptibility at $t=t_3+\twtwo+\twone$, now denoted simply by 
$\chit$, merges into the isothermal curve 
$\tilde{\chi}_{T_1}(\tau_\omega;t)$ from below when the former is 
plotted against $\tilde{t}=t_3+t_{\rm w}^{\rm (eff)}$. Here 
$t_{\rm w}^{\rm (eff)}$ is the effective waiting time given by 
$\twtwo^{z(T_1)/z(T_2)}$ (since $\twtwo \gg \twoneeff$ in this case).
In the case that $\twtwo$ is small enough, we obtain a similar behaviour but 
with $t_{\rm w}^{\rm (eff)} \simeq \twone$ (not shown). This means 
that aging at $T_2$ does not contribute at all to  aging at $T_1$. 
This is nothing but the memory effect and has been observed by many 
experiments.~\cite{VincHOBC,LHOV,Mamiya99} 

In a positive $T$-cycling process with $T_2 > T_1$ shown in 
Fig.~\ref{fig:Tcyc}(b), our simulated data still exhibit the memory 
effect in the sense that the curve at $t_3>0$ lies on the isothermal 
curve of $T_1$ if it is shifted by a proper amount, i.e., 
$t_{\rm w}^{\rm (eff)}$ which depends not only $\twtwo$ but also 
on $\twone$. This type of memory retained over a positive $T$-cycling 
has been indeed observed in the ferromagnetic fine particle 
system.~\cite{Mamiya99} 

\subsection{Continuous $T$-change with an intermittent stop}

We have also performed simulations on a continuous $T$-change with 
an intermittent stop. Following the experimental 
protocol,~\cite{JVHBN} we cool a system by a constant rate from above
$\Tc$ to below $\Tc$ with an intermittent stop at a certain temperature 
$T_{\rm I}$ below $\Tc$, restart the cooling, and then heat it back by 
the same rate as in the cooling. 
\begin{figure}
\leavevmode\epsfxsize=80mm
\epsfbox%
{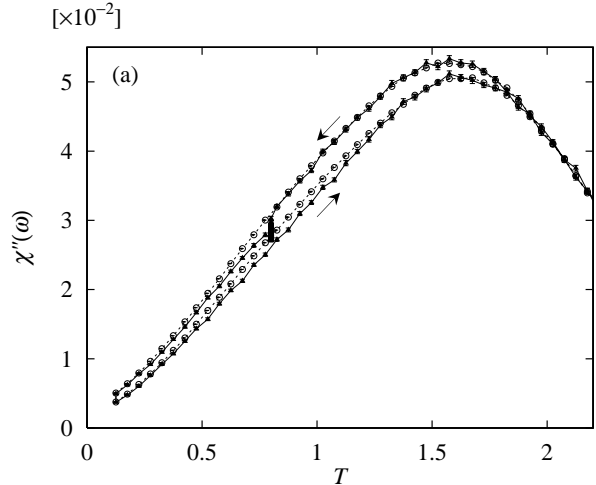}
\caption{$\chit$ in a continuous $T$-change with an intermittent 
stop. We cool and re-heat the system continuously with a constant 
rate of $J/3200$ per MCS from $T=3.0$ to $T=0.1$. The open circles 
represent the reference curves in cooling (upper curve) and 
re-heating (lower curve) protocols. The solid circles represent the same 
protocol except for an intermittent stop in cooling at 
$T_{\rm I}=0.8$ for $32000$ MCS. 
}
\label{fig:conti-T}
\end{figure}
A typical result is shown in Fig.~\ref{fig:conti-T}. In this 
analysis $\chit$ has been directly measured as the response to an ac 
field $h(t)=h_0 \sin(\omega t)$ with $h_0 = 0.3$ and 
$2\pi/\omega = 160$ (MCS).

At the intermittent stop at $T_{\rm I}$, $\chitorg$ relaxes downward 
and deviates from the reference curve of cooling. After restarting 
the cooling, $\chitorg$ is seen to merge into the reference curve. 
It appears that the long time aging at $T_{\rm I}$ does not affect
subsequent cooling at temperatures lower than $T_{\rm I}$ by a 
certain amount. Here we call this phenomenon a rejuvenation-like 
effect, since $\chitorg$ merges with the reference curve but without 
a strong vertical increase in contrast to the rejuvenation seen in the 
experiment.~\cite{JVHBN} By a detailed inspection of
Fig.~\ref{fig:conti-T}, one may notice that the curve $\chitorg$ looks
as if it increases just after the cooling is restarted. We consider
that this behaviour does not indicate the rejuvenation but is due to
the following reason: $\chitorg$ during the stop initially
decreases appreciably but finally its relaxation becomes invisible 
because of its statistical fluctuation. The curve looks as if it 
increases initially at the restart of cooling simply because the
relaxation starts in the middle of the fluctuation. 

When the system is heated back from the lowest temperature of 
cooling process, $\chitorg$ exhibits a dip at around $T=T_{\rm I}$, 
though it is rather shallow. This means that the memory imprinted by 
the long time aging at $T_{\rm I}$ in the cooling process, in fact, 
has not been erased until the system comes back to $T_{\rm I}$ from 
the lower temperatures. To conclude the present SG model exhibits 
the rejuvenation-like and memory effects within a time-window of the 
present simulation. 

\section{Discussions}

We have studied aging phenomena in various $T$-change protocols on 
the 3D Gaussian Ising SG model by Monte Carlo simulations. In 
particular, we have argued that, in a $T$-shift protocol, slow 
aging dynamics after the $T$-change from $T_1$ to $T_2$ at $t_2=0$ 
consists of two regimes: transient ($t_2 \nle \twoneeff$) and 
isothermal ($t_2 \nge \twoneeff$) regimes, where $\twoneeff$ is the 
effective waiting time specified by eq.(\ref{eq:tweff}). In the 
transient regime, re-distribution of thermal weights, from those at 
$T_1$ to $T_2$, of droplet excitations with the characteristic length 
scale $\Ltshf$ is taking place within each domain of the mean size 
$\Rtshf{}$. We called the length scale $\Ltshf$ as the size of 
quasi-domains. 

Suppose, in a negative $T$-shift process, we further decrease 
temperature to $T_3\ (<T_2)$ at $t_2=\twtwo \ (\ll \twoneeff)$ and 
wait a certain period. Then there appear now quasi-domains locally 
equilibrated to $T_3$ within each quasi-domain of $T_2$. Because of
the wide separation of time scales at different length scales 
emphasized in \S 2.1, a nest (or hierarchy) of quasi-domains can be 
realized dynamically when the system is cooled step-wise, or 
continuously. At low temperatures the ac field can only excite 
droplets of the size $L_{T}(1/\omega)$ corresponding to the frequency 
$\omega$. If the size is small enough, a new contribution to the ac 
response will appear every time a new frozen-in domain wall appears 
associated with the quasi-domain due to a step of cooling. 
This picture, {\it quasi-domains within {\rm (}quasi-{\rm )} domains},
is our interpretation of the rejuvenation-like effect seen in 
Fig.~\ref{fig:conti-T}. The picture is somewhat similar to the one
proposed before.~\cite{JVHBN,JNVHB} By our interpretation the
observation  of the rejuvenation-like effect crucially depends on 
time and length scales by which one looks at the system. It is the 
easier to be observed, the shorter and smaller are the time and 
length scales. 

By means of our scenario the memory effect is rather easy to be 
understood. For example, the continuous $T$-change protocol shown in 
Fig.~\ref{fig:conti-T} is interpreted as follows. When the 
system is re-heated from the lowest temperature, the hierarchical 
structure of quasi-domains is erased from the lower levels. 
When the temperature comes back to $T_{\rm I}$, the large domains 
imprinted at the intermittent stop in the cooling process reappear 
and govern the response of the system, which causes the memory 
effect in this protocol. According to our scenario, a memory 
imprinted by aging at a certain temperature $T_{\rm I}$ is erased 
when the system is aged at different temperature, say $T_{\rm d}$, 
for such a long period that the mean size of quasi-domains of $T_d$ 
catches up the domains imprinted at $T_{\rm I}$, as what happens in 
regime 3) of the $T$-shift protocol. Even in that case, the memory 
of the previous thermal history is kept in the effective waiting 
time, i.e., the system looks more aged than the period that it is 
actually aged at $T=T_{\rm d}$ (see eq.(\ref{eq:tc-chi})).

Let us now compare our simulated results with experimental
observations.  As pointed out already in \S 1, in the ferromagnetic 
fine particles system (FFPS), whose aging dynamics has been recently 
studied by Mamiya {\it et al.},~\cite{Mamiya99} the microscopic time, 
$\tau_{\rm mic}$, required for a magnetic moment of each fine 
particle to flip is of the order of $10^{-3}$sec.~\cite{MamiyaPr} 
Therefore the time window of their observation, $1 \sim 10^3$ ksec, 
in unit of $\tau_{\rm mic}$ is not much different from that of our MC 
simulation. It is also noted that in the FFPS each magnetic moment 
behaves as an Ising spin due to strong magnetic anisotropy.

Indeed in the $T$-shift protocol, Mamiya {\it et al.} have observed 
relaxation of $\chish$ which is qualitatively quite similar to our 
simulated $\chiappome$ shown in Fig.~\ref{fig:chiappro}. They have
determined amount of the shift of $\chish$, or $\twoneeff$, by the 
same method that we have used for analysis of $\chiappome$ in 
Fig.~\ref{fig:chiappro}. The effective waiting time they obtained is 
written as $\twoneeff \simeq 3 \times (\twone)^b$ with $b \simeq 1.0$. 
By contrast, $\twoneeff$ extracted from our simulation is given by 
eq.(\ref{eq:tweff}), which, with making use of eq.(\ref{eq:Rt-sim}), 
is reduced to $\twoneeff \propto (\twone)^b$ with  
$b=z(T_2)/z(T_1) \ (\simeq 1.11 \ (1.25)$ for $T_1=0.8$ and
$T_2=0.7 \ (0.6)$). Also aging dynamics of the FFPS in the 
$T$-cycling protocol agrees qualitatively with the results of our 
simulation. In particular, the effective waiting time $\tweff$ which 
depends on $\twone$ (memory effect) is observed even in the positive 
$T$-cycling process, though Mamiya {\it et al.} have claimed that the 
memory is partially lost because some particles with a larger size 
than the average are affected by the chaos effect associated with 
the temperature change. 

For ordinary spin glasses with $\tau_{\rm mic} \sim 10^{-13}$sec, 
there have been only a few experiments, in which $\twoneeff$ in the 
$T$-shift process is systematically studied. Djurberg 
{\it et al.}~\cite{DjuJN} measured the ZFC magnetization of 
Cu:Mn 13.5at\% spin glass with a fixed $\twone$ and varying 
$\Delta T$. Via the FDT similar to eq.(\ref{eq:FDT-chi}) their analysis 
is directly comparable with our simulation on $\Corshzer$ in III. 
The peak position of its relaxation rate with respect to ln$t$ may be 
assigned as $\twoneeff$ (crossover from regime 2) to regime 3)). The 
experimentally observed $\twoneeff$ becomes larger (smaller) than 
$\twone$ in the positive (negative) $T$-shift, which is qualitatively 
consistent with eq.(\ref{eq:tweff}) with monotonic increase of 
$1/z(T)$ on $T$. Quantitatively, however, the shift observed 
experimentally is much larger than what is expected from 
eq.(\ref{eq:z(T)}).

Djurberg {\it et al.}~\cite{DjuJN} furthermore claimed that they 
observed the chaos effect in $T$-shift processes with $|\Delta T|$ 
larger than a certain value. The rejuvenation, or chaos effect in the
$T$-shift and $T$-cycling protocols observed in $\chitorg$ has
been reported for CdCr$_{1.7}$In$_{0.3}$S$_4$~\cite{LHOV,VincBHL} and 
Cu:Mn 2at\%~\cite{AMN93} spin glasses, as already 
mentioned in \S 1. The observed $\chitorg$ is 
claimed~\cite{VincentPr} to indicate that the system is literally 
rejuvenated in the sense described in \S 3.4. Neither such a chaos 
effect nor literal rejuvenation have been seen in our present 
simulations (see also Kisker {\it et al.}~\cite{Kisker-96}). At the 
moment it is not clear yet whether this discrepancy originates simply 
from the difference in time scales of the observations, or from 
something else. 

To conclude, we have simulated various $T$-change protocols of aging 
in the 3D Ising SG model, and have proposed a scenario in 
terms of domains and quasi-domains: the domains ever grow in any aging 
process in the SG phase and the quasi-domains appear within the
domains or within themselves. The results observed in our simulations
are reasonably well interpreted by the scenario. Qualitatively, it is
also consistent with many features observed experimentally,
particularly, those related to the memory effect. For certain
phenomena such as the growth law of $R_T(t)$, the consistency is even
semi-quantitative. Concerned with the rejuvenation, or chaos effect,
however, the scenario may not be applicable, and further studies are
required. 

\section*{Acknowledgments}
We would like to sincerely thank E. Vincent and M. Ocio for their
fruitful discussions and for kindly allowing us to use their data on a
AgMn spin glass. We also thank H. Mamiya for his helpful discussions
on the experiments of his research group. Our thanks are also to 
J.P. Bouchaud, K. Hukushima and P. Nordblad for their useful 
discussions. Two of the present authors (T. K. and H. Y.) were supported 
by Fellowships of Japan Society for the Promotion of Science for 
Japanese Junior Scientists.
This work is supported by a Grant-in-Aid for International Scientific
Research Program, ``Statistical Physics of Fluctuations in Glassy
Systems'' (\#10044064) 
and by a Grant-in-Aid for Scientific Research Program (\#10640362),
from the Ministry of Education, Science and Culture.
The present simulation has been performed on FACOM VPP-500/40 at the 
Supercomputer Center, Institute for Solid State Physics, the University of 
Tokyo.

\vspace*{-3mm}
\bibliography{bibcluster}

\end{document}